\documentclass[twocolumn,showpacs,preprintnumbers,amsmath,amssymb,APSl,prd,nofootinbib,superscriptaddress]{revtex4-1}

\usepackage{dcolumn}% Align table columns on decimal point
\usepackage{bm}% bold math
\usepackage{ifpdf}
\usepackage{hyperref}
\usepackage{dcolumn}
\usepackage{bm}
\usepackage[spanish,english]{babel}
\usepackage{amsfonts}
\usepackage{amssymb}
\usepackage{graphicx}
\usepackage[latin1]{inputenc}

\newcommand{\LL}{\mathcal{L}}

\newcommand{\be}{\begin{equation}}
\newcommand{\en}{\end{equation}}
\newcommand{\bea}{\begin{eqnarray}}
\newcommand{\ena}{\end{eqnarray}}

\begin{document}

\title{Dynamical generation of wormholes with charged fluids in quadratic Palatini gravity}

\author{Francisco S. N.~Lobo}\email{flobo@cii.fc.ul.pt}
\affiliation{Centro de Astronomia e Astrof\'{\i}sica da
Universidade de Lisboa, Campo Grande, Ed. C8 1749-016 Lisboa,
Portugal}
\author{Jesus Martinez-Asencio}
\affiliation{Departamento de F\'{i}sica Aplicada, Facultad de Ciencias, Fase II, Universidad de Alicante, Alicante E-03690, Spain }
\author{Gonzalo J. Olmo} \email{gonzalo.olmo@csic.es}
\affiliation{Departamento de F\'{i}sica Te\'{o}rica and IFIC, Centro Mixto Universidad de
Valencia - CSIC. Universidad de Valencia, Burjassot-46100, Valencia, Spain}
\affiliation{Departamento de F\'isica, Universidade Federal da
Para\'\i ba, 58051-900 Jo\~ao Pessoa, Para\'\i ba, Brazil}
\author{D. Rubiera-Garcia} \email{drubiera@fisica.ufpb.br}
\affiliation{Departamento de F\'isica, Universidade Federal da
Para\'\i ba, 58051-900 Jo\~ao Pessoa, Para\'\i ba, Brazil}


\date{\today}

\begin{abstract}

The dynamical generation of wormholes within an extension of General Relativity (GR) containing (Planck's scale-suppressed) Ricci-squared terms is considered. The theory is formulated assuming the metric and connection to be independent (Palatini formalism) and is probed using a charged null fluid as a matter source. This has the following effect: starting from Minkowski space, when the flux is active the metric becomes a charged Vaidya-type one, and once the flux is switched off the metric settles down into a static configuration such that far from the Planck scale the geometry is virtually indistinguishable from that of the standard Reissner-Nordstr\"om solution of GR. However, the innermost region undergoes significant changes, as the GR singularity is generically replaced by a wormhole structure. Such a structure becomes completely regular for a certain charge-to-mass ratio. Moreover, the nontrivial topology of the wormhole allows to define a charge in terms of lines of force trapped in the topology such that the density of lines flowing across the wormhole throat becomes a universal constant. To the light of our results we comment on the physical significance of curvature divergences in this theory and the topology change issue, which support the view that space-time could have a foam-like microstructure pervaded by wormholes generated by quantum gravitational effects.

\end{abstract}

\maketitle

\section{Introduction}

The Vaidya metric \cite{Vaidya}
\begin{equation}
ds^2=-\left[ 1-\frac{2m(v)}{r} \right] dv^2 +2\epsilon \,dv dr +r^2 d \Omega^2,
\end{equation}
is a nonstatic spherically symmetric solution of the Einstein equations generated by a null stream of radiation. Depending on $\epsilon=+1(-1)$ it corresponds to an ingoing (outgoing) radial flow and $m(v)$ is a monotonically increasing (decreasing) function in the advanced (retarded) time coordinate $-\infty<v<+\infty$. Both the Vaidya solution and its extension to the charged case, the Bonnor-Vaidya solution \cite{BV}, have been widely employed in a variety of physical situations, including the spherically symmetric collapse and the formation of singularities \cite{Lake92}, the study of Hawking radiation and black hole evaporation \cite{radiation}, the gravitational collapse of charged fluids (plasma) \cite{Lasky07} or as a testing tool for various formulations of the cosmic censorship conjecture. In addition to this, several theorems on the existence of exact spherically symmetric dynamical black hole solutions have been established \cite{theor}. In the context of modified gravity, Vaidya-type solutions have been found in metric $f(R)$ gravity coupled to both Maxwell and non-abelian Yang-Mills fields \cite{Ghosh12a} and in Lovelock gravity \cite{Cai08}.

The Vaidya metric has also been used to consider whether a wormhole could be generated out of null fluids. More specifically, in \cite{Hayward02} a crossflow of a two-component radiation was considered, and the resulting solution was interpreted as a wormhole (this analysis extended the results presented in \cite{Gergely02}). It was indeed shown that a black hole could be converted into a wormhole by irradiating the black-hole horizon with pure phantom radiation, which may cause a black hole with two horizons to merge and consequently form a wormhole. Conversely, switching off the radiation causes the wormhole to collapse to a Schwarzschild black hole \cite{Hayward:2001ma}. These results were further extended in \cite{Hayward04} showing that two opposite streams of radiation may support a static traversable wormhole. Furthermore, analytic solutions describing wormhole enlargement were presented, where the amount of enlargement was shown to be controlled by the beaming in and the timing of negative-energy and positive-energy impulses. It was also argued that the wormhole enlargement is not a runaway inflation, but an apparently stable process. The latter issue addressed the important point that though wormholes were possible, and even expected at the Planck scale, macroscopic wormholes were unlikely.

In fact, the generation/construction of wormholes has also been extensively explored in the literature, in different contexts. The late-time cosmic accelerated expansion implies that its large-scale evolution involves a mysterious cosmological dark energy, which may possibly lie in the phantom regime, i.e., the dark energy parameter satisfies $w<-1$ \cite{Planck2}. Now, phantom energy violates the null energy condition, and as this is the fundamental ingredient to sustain traversable wormholes \cite{Morris}, this cosmic fluid presents us with a natural scenario for the existence of these exotic geometries \cite{phantomWH}. Indeed, due to the fact of the accelerating Universe, one may argue that macroscopic wormholes could naturally be grown from the submicroscopic constructions, which envisage transient wormholes at the Planck scale that originally pervaded the quantum foam \cite{Wheeler}, much in the spirit of the inflationary scenario \cite{Roman:1992xj}. It is also interesting to note that self-inflating wormholes were also discovered numerically \cite{Hayward:2004wm}. In the context of dark energy, and in a rather speculative scenario, one may also consider the existence of compact time-dependent dark energy stars/spheres \cite{Lobo:2005uf}, with an evolving dark energy parameter crossing the phantom divide \cite{DeBenedictis:2008qm}. Once in the phantom regime, the null energy condition is violated, which physically implies that the negative radial pressure exceeds the energy density. Therefore, an enormous negative pressure in the center may, in principle, imply a topology change, consequently opening up a tunnel and converting the dark energy star into a wormhole. The criteria for this topology change were also discussed, in particular, a Casimir energy approach involving quasi-local energy difference calculations that may reflect or measure the occurrence of a topology change.

As the Planck scale plays a fundamental importance in quantum gravitational physics, an outstanding question is whether large metric fluctuations may induce a change in topology. Wheeler suggested that at distances below the Planck length, the  metric fluctuations become highly nonlinear and strongly interacting, and thus endow space-time with a foamlike structure \cite{Wheeler}. This behaviour implies that the geometry, and the topology, may be constantly fluctuating, and thus space-time may take on all manners of nontrivial topological structures, such as wormholes. However, paging through the literature, one does encounter a certain amount of criticism to Wheeler's notion of space-time foam, for instance, in that stability considerations may place constraints on the nature or even existence of Planck-scale foamlike structures \cite{Redmount:1992mc}. Indeed, the change in topology of spacelike sections is an extremely problematic issue, and a number of interesting theorems may be found in the literature on the classical evolution of general relativistic space-times \cite{acausal}, namely, citing Visser \cite{Visser}: (i) In causally well-behaved classical space-times the topology of space does not change as a function of time; (ii) In causally ill-behaved classical space-times the topology of space can sometimes change. Nevertheless, researchers in quantum gravity have come to accept the notion of space-time foam, in that this picture leads to topology-changing quantum amplitudes and to interference effects between different space-time topologies \cite{Visser}, although these possibilities have met with some disagreement \cite{dewitt}. Despite the fact that topology-changing processes, such as the creation of wormholes and baby universes, are tightly constrained \cite{Visser:1989ef}, this still allows very interesting geometrical (rather than topological) effects, such as the shrinking of certain regions of space-time to umbilical cords of sufficiently small sizes to effectively mimic a change in topology.

Recently, the possibility that quantum fluctuations induce a topology change, was also explored in the context of Gravity's Rainbow \cite{Garattini:2013pha}. A semi-classical approach was adopted, where the graviton one-loop contribution to a classical energy in a background space-time was computed through a variational approach with Gaussian trial wave functionals \cite{Garattini} (note that the latter approach is very close to the gravitational geon considered by Anderson and Brill \cite{geons4b}, where the relevant difference lies in the averaging procedure). The energy density of the graviton one-loop contribution, or equivalently the background space-time, was then let to evolve, and consequently the classical energy was determined. More specifically, the background metric was fixed to be Minkowskian in the equation governing the quantum fluctuations, which behaves essentially as a backreaction equation, and the quantum fluctuations were let to evolve; the classical energy, which depends on the evolved metric functions, is then evaluated. Analyzing this procedure, a natural ultraviolet (UV) cutoff was obtained, which forbids the presence of an interior space-time region, and may result in a multiply-connected space-time. Thus, in the context of Gravity's Rainbow, this process may be interpreted as a change in topology, and in principle results in the presence of a Planckian wormhole.

In this work, we consider the dynamical generation of wormholes in a quadratic gravity theory depending on the invariants $R=g_{\mu\nu}R^{\mu\nu}$ and $Q=R_{\mu\nu}R^{\mu\nu}$, which are Planck scale-suppressed [see Eq.(\ref{eq:grav-lagrangian}) below for details]. This theory is formulated {\it a la} Palatini, which means that the metric and connection are regarded as independent entities.  Though in the case of General Relativity (GR) this formulation is equivalent to the standard metric approach (where the connection is imposed \emph{a priori} to be given by the Christoffel symbols of the metric) this is not so for modified gravity. Interestingly, the Palatini formulation yields second-order field equations that in vacuum boil down to those of GR and, consequently, are ghost-free, as opposed to the usual shortcomings that plague the metric formulation. To probe the dynamics of our theory, in a series of papers \cite{or12a,lor13} we have studied spherically symmetric black holes with electric charge. As a result we have found electrovacuum solutions that macroscopically are in excellent qualitative agreement with the standard Reissner-Nordstr\"om solution of GR, but undergo important modifications in their innermost structure. Indeed, the GR singularity is generically replaced by a wormhole structure with a throat radius of order $r_c \sim l_P$.  The behaviour of the curvature invariants at $r_c$ shows that for a particular charge-to-mass ratio the space-time is completely regular. The topologically non-trivial character of the wormhole allows us to define the electric charge in terms of lines of electric force trapped in the topology, such that the density of lines of force is given by a universal quantity (independent of the specific amounts of mass and charge). These facts allow to consistently interpret these solutions as geons in Wheeler's sense \cite{Wheeler} and raise the question on the true meaning of curvature divergences in our theory since their existence seems to pose no obstacle for the wormhole extension. Let us note that these wormhole solutions correspond to static solutions of the field equations. Here we shall see that such solutions can be dynamically generated by probing the Minkowski space with a charged null fluid. In this way we obtain a charged Vaidya-type metric such that when the flux is switched off, the space-time settles down into a Reissner-Nordstr\"om-like configuration containing a wormhole structure and thus a multiply-connected topology in its interior. As we shall see, these results have important consequences for the issue of the foam-like structure of space-time. This work largely extends the results and discussion of \cite{lmor14}.

This paper is organized in the following manner: In Sec. \ref{secII}, we present the Palatini formalism for Ricci-squared theories that are used throughout the paper. In Sec. \ref{secIII}, we consider general electrovacuum scenarios with a charged null fluid, and in Sec. \ref{secIV} we solve the gravitational field equations. In Sec. \ref{secV}, we analyze the different contributions to the metric and discuss some particular scenarios. A discussion on the physical implications of these results follows in Sec. \ref{secVI}, where we conclude  with a brief summary and some future perspectives.

\section{Palatini formalism for Ricci-squared theories}\label{secII}

Our initial setup corresponds to that of a generic Palatini Lagrangian coupled to matter, defined by the following action
\begin{equation}\label{eq:action}
S[g,\Gamma,\psi_m]=\frac{1}{2\kappa^2}\int d^4x \sqrt{-g}f(R,Q) +S_m[g,\psi_m]  \ ,
\end{equation}
where $\LL_G=f(R,Q)/(2\kappa^2)$ represents the gravity Lagrangian, $\kappa^2$ is a constant with suitable dimensions (in GR, $\kappa^2 \equiv 8\pi G$), $g_{\mu\nu}$ is the space-time metric,  $R=g^{\mu\nu}R_{\mu\nu}$, $Q=g^{\mu\alpha}g^{\nu\beta}R_{\mu\nu}R_{\alpha\beta}$, $R_{\mu\nu}={R^\rho}_{\mu\rho\nu}$ and
\begin{equation}\label{eq:Riemann}
{R^\alpha}_{\beta\mu\nu}=\partial_{\mu}
\Gamma^{\alpha}_{\nu\beta}-\partial_{\nu}
\Gamma^{\alpha}_{\mu\beta}+\Gamma^{\alpha}_{\mu\lambda}\Gamma^{\lambda}_{\nu\beta}-\Gamma^{\alpha}_{\nu\lambda}\Gamma^{\lambda}_{\mu\beta} \,,
\end{equation}
is the Riemann tensor constructed by the connection $\Gamma \equiv \Gamma^{\lambda}_{\mu\nu}$. The term $S_m[g,\psi_m]$ represents the matter action, where $\psi_m$ are the matter fields, to be specified later.

To obtain the field equations from the action (\ref{eq:action}), in the Palatini approach one assumes that the connection $\Gamma_{\mu\nu}^{\lambda}$, which defines the affine structure, is {\it a priori} independent of the metric, which defines the chrono-geometric structure (see \cite{Zanelli} for a pedagogical discussion). This approaches reduces the number of assumptions on the structure of spacetime beyond GR, and has important consequences for the dynamics of the theory, as we shall see later. The variational principle thus leads to two sets of field equations resulting from the variation of (\ref{eq:action}) with respect to metric and connection as
\begin{eqnarray}
f_R R_{\mu\nu}-\frac{f}{2}g_{\mu\nu}+2f_QR_{\mu\alpha}{R^\alpha}_\nu &=& \kappa^2 T_{\mu\nu}\label{eq:met-varX}\\
\nabla_{\beta}^\Gamma\left[\sqrt{-g}\left(f_R g^{\mu\nu}+2f_Q R^{\mu\nu}\right)\right]&=&0  \ ,
 \label{eq:con-varX}
\end{eqnarray}
respectively. In deriving these field equations, for simplicity, we have set the torsion to zero and assumed $R_{[\mu\nu]}=0$, which guarantees the existence of invariant volumes in our theory \cite{or13a}. The connection equation (\ref{eq:con-varX}) can be solved by means of algebraic manipulations, which are described in a number of previous works \cite{Barragan2010, or12a,lor13}. One thus finds that Eq. (\ref{eq:con-varX}) can be written as
\begin{equation} \label{eq:auxmetric}
\nabla_{\beta}^\Gamma[\sqrt{-h} h^{\mu\nu}]=0 \ ,
\end{equation}
with $h_{\mu\nu}$ defined as
\begin{equation} \label{eq:h-g}
h^{\mu\nu}=\frac{g^{\mu\alpha}{\Sigma_{\alpha}}^\nu}{\sqrt{\det \hat{\Sigma}}} \ , \quad
h_{\mu\nu}=\left(\sqrt{\det \hat{\Sigma}}\right){{\Sigma^{-1}}_{\mu}}^{\alpha}g_{\alpha\nu} \ ,
\end{equation}
where
\begin{equation} \label{eq:Sigmadefinition}
{\Sigma_\alpha}^{\nu}=\left(f_R \delta_{\alpha}^{\nu} +2f_Q {P_\alpha}^{\nu}\right) \ ,
\end{equation}
and ${P_\mu}^\nu\equiv R_{\mu\alpha}g^{\alpha\nu}$. It is easy to verify from Eq.  (\ref{eq:auxmetric}) that $\Gamma_{\mu\nu}^{\lambda}$ can be written  as the Levi-Civita connection of the (auxiliary) metric $h_{\mu\nu}$. It can be shown that $h_{\mu\nu}$ is algebraically related to $g_{\mu\nu}$ and the stress-energy tensor of matter. In fact, in terms of the object ${P_\mu}^\nu$, we can write Eq. (\ref{eq:met-varX}) as
\begin{equation}
f_R {P_\mu}^\nu-\frac{f}{2}{\delta_\mu}^\nu+2f_Q{P_\mu}^\alpha {P_\alpha}^\nu= \kappa^2 {T_\mu}^\nu\label{eq:met-varRQ1} \,,
\end{equation}
or, in matrix form, as (here a hat denotes a matrix)
\begin{equation}
2f_Q\hat{P}^2+f_R \hat{P}-\frac{f}{2}\hat{I} = \kappa^2 \hat{T} \label{eq:met-varRQ2} \ ,
\end{equation}
which represents a quadratic algebraic equation for ${P_\mu}^\nu$ as a function of ${T_\mu}^\nu$. This implies that $R={[\hat{P}]_\mu}^\mu$, $Q={[\hat{P}^2]_\mu}^\mu$, and
${\Sigma_\alpha}^{\nu}$ are just functions of the matter sources.

Using the definition of ${\Sigma_\mu}^\nu$ and the relations (\ref{eq:h-g}), we can write Eq. (\ref{eq:met-varX}) [or, alternatively, Eq. (\ref{eq:met-varRQ1})] as
\begin{equation}
{P_\mu}^\alpha {\Sigma_\alpha}^\nu=R_{\mu\alpha}h^{\alpha\nu} \sqrt{\det \hat\Sigma}=\frac{f}{2}{\delta_\mu^\nu}+\kappa^2{T_\mu}^\nu \,,
\end{equation}
which allows to express the metric field equations using $h_{\mu\nu}$ as follows
\begin{equation} \label{eq:fieldequations}
{R_{\mu}}^{\nu}(h)=\frac{\kappa^2}{\sqrt{\det \hat{\Sigma}}}\left(\LL_G\delta_{\mu}^{\nu}+  {T_\mu}^{\nu} \right) \ .
\end{equation}
This representation of the metric field equations puts forward that $h_{\mu\nu}$ satisfies a set of GR-like second-order field equations. Since $h_{\mu\nu}$ and $g_{\mu\nu}$ are algebraically related, it follows that $g_{\mu\nu}$ also verifies second-order equations. Additionally, we note that in vacuum, $\hat{T}=0$, implies that $\hat{P}$ can be written as $\hat{P}=\Lambda(R^{vac},Q^{vac}_S)\hat{I}$, where the explicit form of $\Lambda(R^{vac},Q^{vac}_S)$ can be found straightforwardly from Eq. (\ref{eq:met-varRQ2}). However, this is not essential for the current discussion. We note that the relations $R^{vac}={P_{\mu}}^{\mu}=4\Lambda(R^{vac},Q^{vac})$, and $Q^{vac}={[P^2]_{\mu}}^{\mu}=4\Lambda^2(R^{vac},Q^{vac})$ imply that the values of $R^{vac}$ and $Q^{vac}$ that simultaneously solve Eq. (\ref{eq:met-varRQ2}) are constant and are related by $Q^{vac}=(R^{vac})^2/4$. In addition, $\hat{P}=\Lambda(R^{vac},Q^{vac})\hat{I}$ also implies that $h_{\mu\nu}$ and $g_{\mu\nu}$ are related by a constant conformal factor [see Eqs. (\ref{eq:h-g}) and (\ref{eq:Sigmadefinition})]. As a result, Eq. (\ref{eq:fieldequations}) tells us that $R_{\mu\nu}(h)=C^{vac}h_{\mu\nu} \ \leftrightarrow \ R_{\mu\nu}(g)=\tilde{C}^{vac}g_{\mu\nu} $, with $C^{vac}$ and $\tilde{C}^{vac}$ constant (and identical in an appropriate system of units). This shows that the vacuum field equations of Palatini theories of the form (\ref{eq:fieldequations}) coincide with the vacuum Einstein equations with a cosmological constant (whose  magnitude depends on the particular gravity Lagrangian $\LL_G$), which is a manifestation of the observed universality of the Einstein equations in the Palatini formalism \cite{Franca}.  These theories, therefore, do not introduce any new propagating degrees of freedom besides the standard massless spin-2 gravitons, and are free from the ghost-like instabilities present in the (higher-derivative) metric formulation of four-dimensional theories containing Ricci-squared terms.

\section{Electrovacuum scenarios with a charged null fluid}\label{secIII}

In this section, we will consider the problem of a spherically symmetric charged space-time perturbed by an ingoing null flux of energy and charge. The electromagnetic field is described by the free Maxwell action plus a coupling to an external current $J^\mu$,
\begin{equation}
S_{em}=-\frac{1}{16\pi} \int d^4x  \sqrt{-g} F_{\mu\nu}F^{\mu\nu}-\int d^4x  \sqrt{-g}A_\mu J^\mu \label{eq:action-em} ,
\end{equation}
where $F_{\mu\nu}=\partial_{\mu}A_{\nu}-\partial_{\nu}A_{\mu}$ is the field strength tensor of the vector potential $A_{\mu}$. The Maxwell stress-energy tensor is obtained as
\begin{equation}
T_{\mu\nu}^{em}=\frac{1}{4\pi}\left[F_{\mu\alpha}{F_{\nu}}^\alpha-\frac{1}{4}F_{\alpha\beta}F^{\alpha\beta} g_{\mu\nu}\right] \label{eq:Tmn-Max0} \ .
\end{equation}
On the other hand, the pressureless flux of ingoing charged matter has a stress-energy tensor
\begin{equation}
T_{\mu\nu}^{flux}= \rho_{in} k_\mu k_\nu   \,,
\label{eq:Tmn-Max1}
\end{equation}
where $k_{\mu}$ is a null vector, satisfying $k_{\mu}k^{\mu}=0$, and $\rho_{in}$ is the energy density of the flux.

In order to write the field equations (\ref{eq:fieldequations}) in combination with the matter source given by Eqs. (\ref{eq:Tmn-Max0}) and (\ref{eq:Tmn-Max1}) in a form amenable to calculations, we need first to obtain the explicit expression of $Q$. To do this we  note that Eq. (\ref{eq:met-varRQ2}) can also be written as
\begin{equation} \label{eq:P-quadratic}
2f_Q\left(\hat{P}+\frac{f_R}{4f_Q}\hat{I}\right)^2=\left(\frac{f}{2}+\frac{f_R^2}{8f_Q}\right)\hat{I}+\kappa^2 \hat{T}   \ .
\end{equation}
In order to obtain an explicit expression for ${P_\mu}^{\nu}$, we need to compute the square root of the right-hand side of this equation. To this effect, let us assume a line element of the form
\begin{equation}\label{eq:ds2g}
ds^2=-A(x,v) e^{2\psi(x,v)}dv^2\pm 2e^{\psi(x,v)}dv dx+r^2(v,x)d\Omega^2 \ ,
\end{equation}
where $-\infty <v< +\infty$ is an Eddington-Finkelstein-like null ingoing coordinate (outgoing if the minus sign is chosen) and $x$ a radial coordinate. Note that $r^2$ is not a coordinate but a function, in general. For this line element we find that a suitable null tetrad is given by
\begin{eqnarray}
k_\mu &=&(-1,0,0,0), \label{eq:tetrad1a}
    \\
l_\mu &=&\left(-\frac{A}{2} e^{2\psi(x,v)},\pm e^{\psi(x,v)},0,0\right) \,, \label{eq:tetrad1b}
\end{eqnarray}
\begin{eqnarray}
m_\mu &=&\left(0,0,\frac{r}{\sqrt{2}},\frac{ir\sin\theta}{\sqrt{2}}\right)\,, \label{eq:tetrad2a}
   \\
\bar{m}_\mu &=&\left(0,0,\frac{r}{\sqrt{2}},-\frac{i r\sin\theta}{\sqrt{2}}\right) \label{eq:tetrad2b}
\end{eqnarray}
and its dual yields
\begin{eqnarray}
k^\mu &=&\left(0,\mp e^{-\psi(x,v)},0,0\right) \,, \label{eq:tetrad3a}
   \\
l^\mu &=&\left(1,\pm \frac{A e^{\psi(x,v)}}{2},0,0 \right) \,, \label{eq:tetrad3b}
\end{eqnarray}
\begin{eqnarray}
m^\mu &=&\left(0,0,\frac{1}{r\sqrt{2}},\frac{i}{\sqrt{2}r\sin\theta}\right) \,, \label{eq:tetrad4a}
   \\
\bar{m}^\mu &=&\left(0,0,\frac{1}{r\sqrt{2}},-\frac{i}{\sqrt{2}r\sin\theta} \right) \,, \label{eq:tetrad4b}
\end{eqnarray}
respectively. Thus, in this representation the only  non-vanishing products are $k^\mu l_\mu=-1$ and $m^\mu \bar{m}_\mu=1$.

\subsection{The matter field equations}

With the above null tetrad, the stress-energy tensor (\ref{eq:Tmn-Max0}) for a spherically symmetric non-null electromagnetic field can be expressed as \cite{Stephani}
\begin{equation} \label{eq:Tem}
T_{\mu\nu}^{em}=\chi \left(m_\mu \bar{m}_\nu+m_\nu \bar{m}_\mu+k_\mu l_\nu+k_\nu l_\mu\right)  \ ,
\end{equation}
where the form of $\chi(v)$ can be obtained by solving explicitly the field equations for the Maxwell field and comparing with Eq. (\ref{eq:Tmn-Max0}) written in matrix representation [see Eq. (\ref{eq:chi}) below].

From the action (\ref{eq:action-em}) and taking into account the presence of the null fluid (\ref{eq:Tmn-Max1}) the Maxwell equations read
\begin{equation}
\nabla_\mu F^{\mu\nu}=4\pi J^\nu \,,
\end{equation}
where $J^\nu\equiv \Omega(v) k^\nu$ is the current of the null ingoing flux, with $\Omega(v)$ a function to be determined.
With this expression and knowing that $\nabla_\mu F^{\mu\nu}\equiv \frac{1}{\sqrt{-g}}\partial_\mu\left(\sqrt{-g} F^{\mu\nu}\right)$, the only non-trivial equations are
\begin{eqnarray}
\partial_x\left(\sqrt{-g} F^{xv}\right)&=&0 \,, \label{eq:em1}\\
\partial_v\left(\sqrt{-g} F^{vx}\right)&=&-4\pi \sqrt{-g} e^{-\psi(x,v)} \Omega  \,, \label{eq:em2}
\end{eqnarray}
From Eq. (\ref{eq:em1}) we find that $r^2 e^{\psi(x,v)}F^{xv}=q(v)$, where $q(v)$ is an integration function. Inserting this back in Eq. (\ref{eq:em2}), it follows that $\Omega(v)=q_v/4\pi r^2$. Note that the function $q_v\equiv \partial_v q(v)$ is our input and, therefore, can be freely specified. Having defined the current that gives consistency to the Maxwell field equations, we can compute explicitly the form of $T_{\mu\nu}^{em}$ in Eq. (\ref{eq:Tem}) to obtain $\chi$, which yields
%$\chi=\frac{q^2(v)}{8\pi r^4}$.
\begin{equation}
\chi=\frac{q^2(v)}{8\pi r^4}\,. \label{eq:chi}
\end{equation}

\subsection{Energy-momentum conservation}

To verify that charge and momentum are conserved in our model we consider the following equation
\begin{equation}
\nabla_\mu {T^\mu}_{\nu, em}= J^\alpha F_{\nu\alpha}+\frac{1}{4\pi}\left[F^{\mu\alpha} \nabla_\mu F_{\nu \alpha}-\frac{1}{2}F^{\alpha\beta}\nabla_\nu F_{\alpha\beta}\right] \ ,\label{eq:Tmnconservation}
\end{equation}
where $\nabla_\mu$ is the derivative operator of the metric $g_{\alpha\beta}$.
Now we use the Bianchi identities $\nabla_{[\nu}F_{\alpha\beta]}=0$ to express
\begin{equation}
F^{\alpha\beta}\nabla_\nu F_{\alpha\beta}=-F^{\alpha\beta}(\nabla_\beta F_{\nu\alpha}+\nabla_\alpha F_{\beta\nu})=2F^{\beta\alpha}\nabla_\beta F_{\nu\alpha} \ .
\end{equation}
Inserting this result in Eq. (\ref{eq:Tmnconservation}) we find $\nabla_\mu {T^\mu}_{\nu, em}=J^{\alpha} F_{\nu\alpha}$.On the other hand, the null fluid yields
\begin{equation}
\nabla_\mu {T^{\mu}}_{\nu, \ flux}=k_\nu\nabla_\mu \left(\rho_{in}k^\mu\right)+\rho_{in}k^\mu\nabla_\mu k_\nu \ .
\end{equation}
One can see by direct calculation that $k^\mu\nabla_\mu k_\nu=0$, which verifies that $k_\mu$ is a geodesic vector.  Since $\nabla_\mu {T^{\mu}}_{\nu, \ Total}=\nabla_\mu {T^{\mu}}_{\nu,em} + \nabla_\mu {T^{\mu}}_{\nu,flux}=0$, contracting with $l_\nu$ we find
\begin{equation}
\partial_\mu \left(\sqrt{-g}\rho_{in}k^\mu\right)+\sqrt{-g}F^{\mu\nu} l_\mu J_\nu=0  \ ,
\end{equation}
which becomes
\begin{equation}\label{eq:conservation}
\partial_x\left(\rho_{in}r^2\right)=e^{\psi(x,v)} \frac{q q_v}{4\pi r^2} \ .
\end{equation}
This condition should be satisfied by the solution of the problem on consistency grounds. We note that an analogous procedure for the derivation of these expressions can be carried out when a magnetic field, and consequently, a magnetic flux, is present. Thereby, the expressions in the previous subsections can be trivially extended to those of the magnetic case by just swapping the electric charge $q$ with a magnetic charge $g$.

\subsection{The ${\Sigma_{\mu}}^{\nu}$ matrix.}

With the representation of the stress-energy tensor given by Eq. (\ref{eq:Tem}), we can proceed to obtain the square root of Eq. (\ref{eq:P-quadratic}) for the ingoing flux of charged null matter represented by the stress-energy tensor in Eq. (\ref{eq:Tmn-Max1}). To do this, we identify the right-hand side of Eq. (\ref{eq:P-quadratic}) with the squared matrix
\begin{eqnarray} \label{eq:M}
{M_a}^c M{_c}^b &=& \lambda \delta_a^b + \kappa^2\chi (m_a\bar{m}^b + m^b \bar{m}_a +k_a l^b+k^b l_a) \nonumber \\ &&+ \kappa^2 \rho_{in} k_a k^b \ ,
\end{eqnarray}
where $\lambda\equiv\frac{f}{2} + \frac{f_R^2}{8f_Q}$. We now propose the ansatz
\begin{eqnarray} \label{eq:ansatzM}
{M_a}^b&=&\alpha \delta_a^b + \beta (m_a \bar{m}_b + m^b \bar{m}_a)+\gamma (k_al^b+k^bl_a) \nonumber \\ && + \delta k_a k^b + \epsilon l_al^b,
\end{eqnarray}
where $\alpha, \beta, \gamma, \delta, \epsilon$ are functions to be determined by matching the right-hand side of Eq. (\ref{eq:M}) with the square ${M_a}^c M{_c}^b$ using the ansatz  (\ref{eq:ansatzM}). This leads to the set of equations
\begin{eqnarray}
\alpha^2 = \lambda ; \quad  \beta(\beta+2\alpha)&=&\kappa^2 \chi ; \quad  \gamma(2\alpha-\gamma)-\delta \epsilon = \kappa^2 \chi \nonumber \\
2\delta(\alpha-\gamma)= \kappa^2 \rho_{in} ; && \quad  2\epsilon(\alpha-\gamma)=0
\end{eqnarray}
whose unique consistent solution is
\begin{eqnarray}
&&\alpha = \lambda^{1/2} \,;\qquad \beta =-\lambda^{1/2} + \sqrt{\lambda + \kappa^2 \chi} \nonumber \\
&& \hspace{-0.4cm}  \gamma =\lambda^{1/2}-\sqrt{\lambda - \kappa^2 \chi} \,;\quad \delta=\frac{\kappa^2 \rho_{in}}{2\sqrt{\lambda - \kappa^2 \chi}} \,;\quad  \epsilon=0  \,.  \label{eq:abgde}
\end{eqnarray}
With these results, we can use the expression
\begin{equation}
\hat{M}=\sqrt{2f_Q} \left(\hat{P} + \frac{f_R}{4f_Q} \hat{I} \right) \,,
\end{equation}
to write the matrix ${\Sigma_{\mu}}^{\nu}$ defined in Eq. (\ref{eq:Sigmadefinition}) as
\begin{equation}
{\Sigma_{\mu}}^{\nu}=\frac{f_R}{2}{\delta_{\mu}}^{\nu} + \sqrt{2f_Q}{M_{\mu}}^{\nu} \,,
\end{equation}
which finally becomes
\begin{eqnarray}
{\Sigma_{\mu}}^{\nu}&=& \left(\frac{f_R}{2}+ \sqrt{2f_Q} \lambda^{1/2} \right)\delta_{\mu}^{\nu}
\nonumber \\
&&+ \sqrt{2f_Q} \Big[\beta (m_{\mu} \bar{m}^{\nu} +m^{\nu} \bar{m}_{\mu})
\nonumber \\
&& + \gamma(k_{\mu}l^{\nu}+ k^{\nu} l_{\mu}) + \delta k_{\mu}k^{\nu}\Big] \,.
\end{eqnarray}
Note that this expression only depends on the electromagnetic, $\chi$, and fluid, $\rho_{in}$, parameters and on the Lagrangian density $f(R,Q)$.

\subsection{The $f(R,Q)$ Lagrangian}

In what follows we shall only be concerned with the quadratic Lagrangian
\begin{equation} \label{eq:grav-lagrangian}
f(R,Q)=R+l_P^2 (a R^2+bQ) \ ,
\end{equation}
where $l_P^2 \equiv \hbar G/c^3$ represents Planck's length squared; $a$ and $b$ are dimensionless constants. The physical reason underlying this models is the fact that quadratic corrections of the form above arise in the quantization of fields in curved space-time \cite{PT} and also in the low-energy limits of string theories \cite{strings1,Ortin}. It has also been argued that this kind of models are natural in an effective field theory approach to quantum gravity \cite{Cembranos}. Palatini Lagrangians with quadratic and higher-order curvature corrections also arise in effective descriptions \cite{o08} of the dynamics of loop quantum cosmology \cite{lqc}, a scenario in which the big bang singularity is replaced by a cosmic bounce.  Alternatively, one could use other models of Palatini gravity, such as the proposal of Deser and Gibbons \cite{Deser}, dubbed Eddington-inspired Born-Infeld gravity \cite{Banados}, which in the static electrovacuum case is exactly equivalent (not only perturbatively) to the quadratic Lagrangian (\ref{eq:grav-lagrangian}), as shown in \cite{orh14}. As a working hypothesis, we shall assume that neither the quadratic Lagrangian (\ref{eq:grav-lagrangian}), nor the perturbation induced by the flux will spoil the geometrical nature of gravitation as we approach the scale where the $l_P^2$ effects in (\ref{eq:grav-lagrangian}) begin to play an important role.

Tracing in Eq. (\ref{eq:met-varX}) with the metric $g^{\mu\nu}$ it follows that $R=-k^2 T$, where $T$ is the trace of the stress-energy tensor. For the electric field  we are considering, one has $T=0$, which implies $R=0$. As a result, the dependence of the Lagrangian on the parameter $a$ becomes irrelevant. Note, however, that for nonlinear theories of electrodynamics (for which $T \neq 0$), the parameter $a$ does play a role \cite{or13b,or13c}. From now on we consider the case $b>0$ and, for simplicity, set $b=1$. To obtain the expression for $Q$ we take the trace of ${M_a}^{b}$ and use the tetrad relations to write
\begin{equation} \label{eq:forQ}
\frac{1}{\sqrt{2}l_P}=\sqrt{\lambda+\kappa^2 \chi} - \sqrt{\lambda-\kappa^2 \chi} \,.
\end{equation}
For our theory given by Eq. (\ref{eq:grav-lagrangian}), we have $f=l_P^2 Q$, $f_R=1, f_Q=l_P^2$ and then from Eq. (\ref{eq:Tmn-Max0}) we obtain $\lambda=\frac{1}{8l_P^2}(1+4l_P^2 Q)$. Inserting this in Eq. (\ref{eq:forQ}), we obtain the solution
\begin{equation}
Q=4\kappa^4 \chi^{2}=\frac{\tilde{\kappa}^4q^4}{r^8} \,,
\end{equation}
where $\tilde{\kappa}^2=\kappa^2/(4\pi)$. We note that this expression remains unchanged if the null fluid is absent \cite{or12d}.

To obtain the field equations (\ref{eq:fieldequations}) for the theory given by Eq. (\ref{eq:grav-lagrangian}) we need both the explicit expression of ${\Sigma_{\mu}}^{\nu}$ and of $(\LL_G\delta_{\mu}^{\nu}+ {T_{\mu}}^{\nu})$ appearing on the right-hand side of Eq. (\ref{eq:fieldequations}). From the tetrad definitions (\ref{eq:tetrad1a})-(\ref{eq:tetrad4b}) it is easily seen that
\begin{eqnarray} \label{eq:m}
m_{\mu}\bar{m}^{\nu}+m^{\nu}\bar{m}_{\mu}=\left(
\begin{array}{cc}
\hat{0}& \hat{0} \\
\hat{0} & \hat{I} \\
\end{array}
\right) \,,
\\
\hspace{0.1cm}
k_{\mu}l^{\nu}+l^{\nu}k_{\mu}=\left(
\begin{array}{cc}
-\hat{I}& \hat{0} \\
\hat{0} & \hat{0} \\
\end{array}
\right)\,,
\end{eqnarray}
where $\hat{I}$ and $\hat{0}$ are the $2 \times 2$ identity and zero matrices, respectively. In this formalism, we immediately recover the usual expression for a spherically symmetric electromagnetic field, namely, $T_{\mu\nu}^{em}=\chi \, \text{diag}(-1,-1,1,1)$. On the other hand, for the null fluid contribution we have
\begin{equation} \label{eq:k}
k_{\mu}k^{\nu}=\left(
\begin{array}{cccc}
0 & e^{-\psi} & 0 & 0 \\
0 & 0 & 0 & 0 \\
0 & 0 & 0 & 0 \\
0 & 0 & 0 & 0  \\
\end{array}
\right) \ .
\end{equation}
Taking into account all these elements, one readily finds that
\begin{equation} \label{eq:Sigma}
{\Sigma_{\mu}}^{\nu}=\left(
\begin{array}{cccc}
\sigma_-  & \sigma_{in} & 0 & 0 \\
0 & \sigma_- & 0 & 0 \\
0 & 0 & \sigma_+ & 0 \\
0 & 0 & 0 & \sigma_+  \\
\end{array}
\right) ,
\end{equation}
where
\begin{eqnarray}\label{eq:sigmapm}
\sigma_{\pm}&=&1\pm \frac{\tilde{\kappa}^2 l_P^2 q^2(v)}{r^4} \,, \\
\sigma_{in}&=&\frac{2\kappa^2 l_P^2 \rho_{in}}{1-2\tilde{\kappa}^2 l_P^2 q^2(v)/r^4} \,. \label{eq:sigmain}
\end{eqnarray}
 With all these results, we can finally write the field equations (\ref{eq:fieldequations}) for our problem as
\begin{equation}\label{eq:Rmn}
{R_{\mu}}^{\nu}(h)=\left(
\begin{array}{cccc}
-\frac{\tilde{\kappa}^2q^2(v)}{2r^4\sigma_+} & \frac{e^{-\psi}\kappa^2\rho_{in}}{\sigma_+\sigma_-}  & 0 & 0  \\
0 & -\frac{\tilde{\kappa}^2q^2(v)}{2r^4\sigma_+} & 0 & 0  \\
0& 0& \frac{\tilde{\kappa}^2q^2(v)}{2r^4\sigma_-} & 0 \\
0& 0& 0 & \frac{\tilde{\kappa}^2q^2(v)}{2r^4\sigma_-}
\end{array}
\right)  .
\end{equation}
Note that in the limit $l_P \rightarrow 0$, we have $\sigma_{\pm} \rightarrow 1$ and $\sigma_{in}\to 0$, which entails $h_{\mu\nu}=g_{\mu\nu}$ and Eq. (\ref{eq:Rmn}) recovers the equations of GR.

\section{Solving the field equations}\label{secIV}

Having obtained the field equations in the form of Eq. (\ref{eq:Rmn}), we now proceed to solve them as follows. Firstly, we propose a spherically symmetric line element associated to the metric $h_{\mu\nu}$ following the structure given in Eq. (\ref{eq:ds2g}), namely
\begin{eqnarray}
d\tilde{s}^2&=& -F(v,x)e^{2\xi(v,x)}dv^2+2e^{\xi(v,x)}dv dx  \nonumber\\
&&+ \tilde{r}^2(v,x)d\Omega^2 \,. \label{eq:ds2h}
\end{eqnarray}
Direct comparison of $g_{\mu\nu}$ and $h_{\mu\nu}$ using $\hat{\Sigma}$ implies that
\begin{eqnarray}
g_{vv}&=&\frac{h_{vv}}{\sigma_+}+\frac{\sigma_{in} h_{vx}}{\sigma_+\sigma_-} \,, \label{eq:h-to-g} \\
g_{vx}&=& \frac{h_{vx}}{\sigma_+}  \label{eq:gvx} \,,
\end{eqnarray}
which leads to $e^\psi=\frac{e^\xi}{\sigma_+}$ and $\tilde{r}^2=r^2\sigma_-$.
Note that given the $v$-dependence of $q(v)$ and the $(x,v)$-dependence of $\sigma_\pm$ in Eq. (\ref{eq:sigmapm}), it is reasonable to expect {\it a priori} some $v$-dependence on $\tilde{r}$ [as we have assumed in Eq. (\ref{eq:ds2h})]. In fact, from the relation $\tilde{r}^2=r^2\sigma_-$, one finds
\begin{equation}\label{eq:rtr}
r^2=\frac{\tilde{r}^2+\sqrt{\tilde{r}^4+4l_P^2 \tilde{\kappa}^2q^2(v)}}{2} \ ,
\end{equation}
which establishes a non-trivial relation between $r,\tilde{r}$, and $q(v)$. For this reason, we have not used $\tilde{r}$ as a variable and have kept the independent coordinate $x$ in the  non-spherical sector of the line element (\ref{eq:ds2h}). The explicit relation between $\tilde{r}, x$ and $v$ must thus follow from the field equations.

From the line element (\ref{eq:ds2h}) we obtain, using the algebraic manipulation package xAct \cite{JMMG},
\begin{equation}
{R_x}^v\equiv\frac{2e^{-\xi}(\tilde{r}_x \xi_x-\tilde{r}_{xx})}{\tilde{r}}=0 \ .
\end{equation}
This implies that $e^{\xi(x,v)}=C(v)\tilde{r}_x$, and inserting the latter in Eq. (\ref{eq:ds2h}), yields
\begin{equation} \label{eq:ds2hb}
d\tilde{s}^2= -F(v,x)\tilde{r}_x^2dv^2+2\tilde{r}_xdv dx+\tilde{r}^2(v,x)d\Omega^2  \ .
\end{equation}
where the function $C(v)$ has been reabsorbed into a redefinition of $v$. Working now with the ansatz (\ref{eq:ds2hb}), we get
  \begin{equation}
{R_x}^v\equiv-\frac{2\tilde{r}_{xx}}{\tilde{r}}=0 \ ,
\end{equation}
which implies that
  \begin{equation}
\tilde{r}=\alpha(v)x+\beta(v) \ ,
\end{equation}
where $\alpha(v)$ and $\beta(v)$ are, so far, two arbitrary functions.

Assuming that $F(x,v)=1-2M(x,v)/\tilde{r}$, we obtain
\begin{eqnarray}
R_{vx}&\equiv&\frac{1}{\tilde{r}}\left[M_{xx}-2\alpha_v\right]={R_v}^v={R_x}^x \,,
   \\
{R_\theta}^\theta &\equiv& \frac{1}{\tilde{r}^2}\left[1-\alpha^2-2\beta\alpha_v-2\alpha[\beta_v+(2x\alpha_v-M_x)]\right] \,,
    \\
R_{vv}&\equiv&-\frac{1}{\tilde{r}^2}\left[(\tilde{r}-2M)M_{xx}+2\tilde{r}\tilde{r}_{vv}
+2\tilde{r}_vM_x-2\alpha M_v\right] \,.
\end{eqnarray}
From the first of these equations, we find
\begin{equation}\label{eq:Mx}
M_x=\frac{\tilde{\kappa}^2q^2}{4\alpha r^2}+2\alpha_v x \ .
\end{equation}
 Inserting this result in ${R_\theta}^\theta$ and performing some manipulations, one obtains
\begin{equation}
1-\alpha^2=\partial_v(\alpha\beta) \ .
\end{equation}
A consistent solution of this equation is $\beta=0$, and $\alpha=1$, which implies that $\tilde{r}=x$ is independent of $v$. Assuming this solution from now on, we find that
\begin{equation}
{R_v}^x\equiv R_{vv}+FR_{vx}=\frac{2M_v}{\tilde{r}^2} \ .
\end{equation}
Since ${R_v}^x=\frac{\kappa^2\rho_{in}}{\sigma_-}$, the above relation implies
\begin{equation}\label{eq:Mv}
M_v=\frac{\kappa^2\rho_{in} r^2}{2} \,.
\end{equation}
With the above results, and using the relations $\partial_v(q^2/r^2)\big|_{x}=2q q_v/r^2\sigma_+$ (at constant $x$) and $dr\big|_v=\frac{\sigma_{-}^{1/2}}{\sigma_{+}}dx$ (at constant $v$), one can show that the integrability condition $\partial_v M_x=\partial_x M_v$ implies the conservation equation (\ref{eq:conservation})  (where the relation (\ref{eq:rtr}), which leads to $\sqrt{x^4+4l_P^2 \tilde{\kappa}^2q^2}=r^2\sigma_+$, must be used).

From Eq. (\ref{eq:Mx}), with $\alpha=1$, by direct integration we find
\begin{equation}
M(x,v)=\frac{\tilde{\kappa}^2q^2(v)}{4}\int \frac{dx}{r^2}+\gamma(v) \ .
\end{equation}
Computing $M_v$ from this expression and comparing with Eq. (\ref{eq:Mv}), we find (recall that  $\partial_v(q^2/r^2)\big|_{x}=2q q_v/r^2\sigma_+$)
\begin{equation}
\gamma_v=\frac{{\kappa}^2}{2}\left(\rho_{in} r^2-\frac{q q_v}{4\pi}\int \frac{dx}{r^2\sigma_+}\right) \ .
\end{equation}
Since $\gamma=\gamma(v)$, defining $L(v) \equiv \gamma_v$ as the \emph{luminosity function}, it follows that
\begin{equation}
\rho_{in} r^2=\frac{2}{\kappa^2} \left[L(v) + \frac{\kappa^2 q q_v}{8\pi}\int \frac{dx}{r^2\sigma_+} \right] \,, \label{eq:constraint}
\end{equation}
which is fully consistent with the conservation equation (\ref{eq:conservation})  because Eq. (\ref{eq:gvx}) and the subsequent manipulations imply $e^\psi=1/\sigma_+$.  In summary, we conclude that
\begin{eqnarray}
\tilde{r}(x,v)&=&x \ , \\
r^2(x,v)&=&\frac{x^2+\sqrt{x^4+4l_P^2 \tilde{\kappa}^2q^2(v)}}{2}  \label{eq:rtx0} \ , \\
F(x,v)&=&1-\frac{2M(x,v)}{x} \label{eq:mass-1} \,, \\
M(x,v)&=& M_0+\gamma(v)+\frac{\tilde{\kappa}^2q^2(v)}{4}\int \frac{dx}{r^2} \,, \label{eq:Mass}\\
\gamma(v)&=& \int dv L(v)  \,,\\
\rho_{in}&=&\frac{2}{\kappa^2 r^2} \left[L(v) + \frac{\kappa^2 q q_v}{8\pi}\int \frac{dx}{r^2\sigma_+} \right]  \label{eq:rhoin} \,.
\end{eqnarray}
This set of equations provides a consistent solution to the Palatini $f(R,Q)$ Lagrangian (\ref{eq:grav-lagrangian}) with null and non-null electromagnetic fields satisfying $\nabla_\mu F^{\mu\nu}=\frac{q_v}{4\pi r^2}k^\nu$, where $q_v\equiv \partial_v q(v)$ and $\gamma_v\equiv L(v)$ are free functions. Their dependence on $v$ reflects the presence of the charged stream of null particles.

Given the structure of the mass function in Eq. (\ref{eq:Mass}) and to make contact with previous results on static configurations, we find it useful to write it as
\begin{equation}
M(x,v)= M_0+\gamma(v) +\frac{r_q(v)^2}{4r_c(v)}\left(\int dz G_z\right)\Big|_{z=r/r_c} \ ,
\end{equation}
with $r_c(v)=\sqrt{r_q(v)l_P}$, $z(x,v)=r(x,v)/r_c(v)$ and
\begin{equation}
G_z=\frac{z^4+1}{z^4 \sqrt{z^4-1}} \label{eq:Gz} \,,
\end{equation}
where we have used the relation $dr/dx=\sigma_{-}^{1/2}/\sigma_{+}$ (at constant $v$). This can be expressed in a more compact form as
\begin{equation}
M(x,v)=M(v)\left[1+\delta_1(v) G\left(z\right)\right]\big|_{z=\frac{r}{r_c}}\ ,
\end{equation}
where $M(v)=M_0+\gamma(v)\equiv r_S(v)/2 $ and
\begin{equation} \label{eq:delta1}
\delta_1(v)=\frac{1}{2r_S(v)} \sqrt{\frac{r_q^3(v)}{l_P}} \ .
\end{equation}

The function $G(z)$ can be written as an infinite power series and its form was given in \cite{or12a}.
Using these results,  we can write $g_{vv}$ in Eq. (\ref{eq:h-to-g}) as
\begin{equation} \label{eq:metric-ingoing}
g_{vv}=-\frac{F(x,v)}{\sigma_{+}} + \frac{2l_P^2 \kappa^2 \rho_{in}}{\sigma_{-}(1-\frac{2r_c^4}{r^4})} \,,
\end{equation}
where  (recall that $z=r(x,v)/r_c(v)$)
\begin{equation}
F(x,v)=1-\frac{1+\delta_1 (v) G(z)}{\delta_2(v) z \sigma_{-}^{1/2}} \label{eq:Ffunction}  \,,
\end{equation}
and we have introduced the parameter
\begin{equation} \label{eq:delta2}
\delta_2(v)=\frac{r_c(v)}{r_S(v)} \,.
\end{equation}
Using the above results, the line element (\ref{eq:ds2g}) becomes
\begin{eqnarray}
ds^2&=&-\left[\frac{1}{\sigma_+}\left(1-\frac{1+\delta_1 (v) G(z)}{\delta_2(v) z \sigma_{-}^{1/2}}\right)- \frac{2l_P^2 \kappa^2 \rho_{in}}{\sigma_{-}(1-\frac{2r_c^4}{r^4})}\right]dv^2
    \nonumber\\
 &&+ \frac{2}{\sigma_+}dvdx+r^2(x,v) d\Omega^2 \ , \label{eq:ds2final}
\end{eqnarray}
%where $\sigma_+=\sigma_+(x,v)$ is specified by (\ref{eq:sigmapm}) together with (\ref{eq:rtr}) taking $\tilde{r}=x$.
Equation (\ref{eq:ds2final}) with the definitions given by Eqs. (\ref{eq:sigmapm}), (\ref{eq:delta1}) and (\ref{eq:delta2}), and the function $G_z$ in Eq. (\ref{eq:Gz}), which contains the contribution of the non-null electromagnetic field, constitutes the main result of this paper.

\section{Physical properties}\label{secV}

In the previous section, we found the exact analytical solution to the problem of a spherically symmetric ingoing null fluid carrying electric charge and energy in a space-time whose dynamics is governed by the Palatini theory given by Eq. (\ref{eq:grav-lagrangian}). In this section, we discuss the different contributions appearing in the line element (\ref{eq:ds2final}) and their properties.

\subsection{The GR limit}

Let us first note that when $l_P \rightarrow 0$ in the Lagrangian density (\ref{eq:grav-lagrangian}) we recover the GR limit, since it implies $\sigma_{\pm} \rightarrow 1$ and allows to perform the integration in $dx=dr \sigma_{+}/\sigma_{-}^{1/2}=dr$, finding that $\int dx/r^2=-1/r$, which leads to  $G(z) = -1/z$. With these elements the metric component $g_{vv}$ in Eq. (\ref{eq:metric-ingoing}) boils down to
\begin{equation}
g_{vv}=-\left(1 - \frac{r_s(v)}{r} + \frac{r_q(v)^2}{2r^2} \right) \,,
\end{equation}
while the relation (\ref{eq:constraint}) becomes
\begin{equation}
\rho_{in}r^2=\frac{2}{\kappa^2} \left( L(v) - \frac{\kappa^2 q q_v}{8\pi r} \right) \,.
\end{equation}
Thus, these expressions reproduce the well known Bonnor-Vaidya solution of GR \cite{BV}.

\subsection{Uncharged solutions with null fluid}

This dynamical scenario was considered in \cite{or12d}, where more details and examples can be found. Here we summarize the main features of this case.  When the electrovacuum field is not present ($q=0$, $\sigma_{\pm} \rightarrow 1$),  the line element (\ref{eq:ds2final}) becomes \cite{or12d}
\begin{equation} \label{eq:ln-ud}
ds^2=-B(v,r)dv^2+2dvdr + r^2 d \Omega^2 \,,
\end{equation}
with
\begin{equation} \label{eq:bf-ud}
B(v,r)=1-\frac{2 M(v)}{r} - \frac{Q^2(v)}{r^2} \,,
\end{equation}
where $M(v)\equiv\int_{v_0}^v L(v')dv'$ represents the mass term, $Q^2(v)\equiv 4L(v)/\rho_P$ represents a charge-like term, and we have defined $\rho_P=\frac{c^2}{l_P^2 G} \sim 10^{96}kg/m^3$ as the Planck density. The \emph{luminosity} function $L(v)=\kappa^2 r^2 \rho_{in}/2$ follows from Eq. (\ref{eq:Mv}). The metric (\ref{eq:ln-ud}) is formally that of a (nonrotating) Reissner-Nordstr\"om black hole but with the wrong sign in front of the charge term. The single horizon of this solution is located at $r_{+}(v)=M(v) + \sqrt{M(v)+Q^2(v)}$  and is larger than in the dynamical Schwarzschild solution of GR, $r_S(v)=2M(v)$.

When the flux of radiation ceases, $Q^2(v)$ vanishes and $r_+$  retracts to its GR value $r_S$. The metric function (\ref{eq:bf-ud}) puts forward that the  null fluid is leaving its imprint on the structure of the space-time not only through its integrated luminosity (the mass term $M(v)$), but also directly through the luminosity function $L(v)$ (suppressed by the Planck density), which contains full details about the distribution of the incoming fluid. It should be noted that if a scalar field is quantized in such a background, the field mode functions will be sensitive to $L(v)$, thus having access to all the information contained in the source that forms the black hole. As a result, the emitted Hawking quanta will contain crucial information not only about the integrated energy profile $M(v)$, but also about the most minute details of its time distribution $L(v)$.

\subsection{Static charged configurations} \label{sec:static}

When there is no incoming flux of charge and energy, $q(v)$ and $M(v)$ remain constant. In this case, the metric gets simplified in a number of ways. Firstly, the term $\rho_{in}$ disappears, and $\delta_1(v)$, $\delta_2(v)$, and $r_c(v)$ take the constant values $\delta_1^{(0)}$, $\delta_2^{(0)}$, and $r_c^{(0)}$, respectively, where the superindex denotes the amounts of mass and charge, $M=M_0$ and $q=q_0$, characterizing the solution.  The line element (\ref{eq:ds2final}) can then be written as
\begin{eqnarray}
ds^2&=&-\frac{\left(1-\frac{1+\delta_1^{(0)} G(z)}{\delta_2^{(0)} z \sigma_{-}^{1/2}}\right)}{\sigma_+(x)}dv^2+\frac{2dvdx}{\sigma_+(x)}+r^2(x)d\Omega^2 \ , \label{eq:ds2static1}
\end{eqnarray}
where here $r=r_c^{(0)}z(x)$ is just a function of $x$, i.e., there is no time-dependence on $v$. Accordingly, $\sigma_\pm=\sigma_\pm(z)$ and $G(z)$ are $v$-independent functions. For $|x|\gg r_c^{(0)}$, one finds that $r^2(x)\approx x^2$, $\sigma_\pm\approx 1$, and $g_{vv}\approx -\left(1-\frac{r_S^{(0)}}{r}+\frac{{r_q^{(0)}}^2}{2r^2}\right)$, which turns (\ref{eq:ds2static1}) into the expected GR limit.

\begin{figure}[h]
\includegraphics[width=0.45\textwidth]{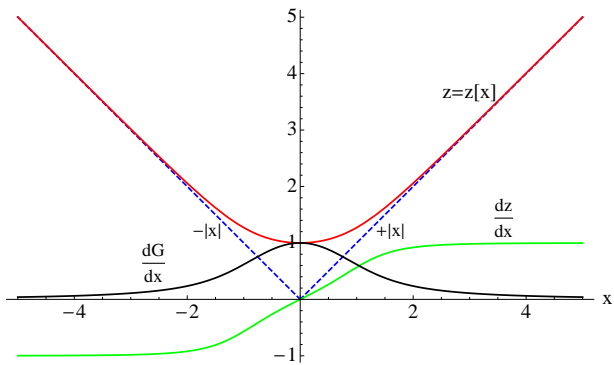}
\caption{The minimum of the radial function $z(x)$ implies the existence of a wormhole extension of the geometry, with $x$ covering the whole real axis  $-\infty <x<+\infty$. Note the smoothness of the function $dG/dx$ and the bounce of $z(x)$ at $x=0$. In this plot, $r_c=1$.
\label{fig:WH_extension}}
\end{figure}

On the other hand, from the relation (\ref{eq:rtx0}), it is easy to see that $r(x)$ reaches a minimum $r_{min}=r_c^{(0)}$ at $x=0$.
At that point, one can verify \cite{or12a} that curvature scalars generically diverge except if the charge-to-mass ratio $\delta_1^{(0)}$ takes the value $\delta_1^{(0)}=\delta_1^*$ \footnote{In terms of horizons, configurations with $\delta_1>\delta_1^*$ are similar to the standard Reissner-Norstr\"om solution of GR, having two horizons, a single (extreme) one or none, while those with $\delta_1<\delta_1^*$  have always a single (non-degenerate) horizon, resembling the Schwarzschild solution. For more details see the first of Refs.\cite{or12a}.}, where $\delta_1^* \simeq 0.572$ is a constant that appears in the series expansion of $G(z)=-1/\delta_1^*+2\sqrt{z-1}+\ldots$ as $z\to 1$. The smoothness of the geometry when $\delta_1^{(0)}=\delta_1^*$ together with the fact that $r(x)$ reaches a minimum at $x=0$ allow to naturally extend the coordinate $x$ to the negative real axis, thus showing that the radial function $r^2(x)$ bounces off to infinity as $x\to -\infty$ (see Fig.\ref{fig:WH_extension}).
This implies the existence of a wormhole structure with its throat located at $r=r_c^{(0)}$, where $dr/dx=0$. This interpretation is further supported by the existence of an electric flux coming out from the wormhole mouth and responsible for the spherically symmetric electric field. An electric field of this kind does not require the existence of point-like sources for its generation, as first shown by Wheeler and Misner in \cite{Misner:1957mt}. The non-trivial wormhole topology implies that the flux $\Phi= \int_S *F$, where $*F$ is the 2-form dual to the Faraday tensor, through any closed 2-surface $S$ enclosing one of the wormhole mouths is non-zero and can be used to define a charge $\Phi=4\pi q$. On practical grounds, there is no difference between this kind of charge, arising from a pure electric field trapped in the topology (going through a wormhole), and a standard point-like charge. Remarkably, one can easily verify \cite{or12a} that this flux is independent of the particular value of $\delta_1^*$, which entails that the wormhole structure exists even when the curvature scalars diverge at $r=r_c^{(0)}$. This result gives consistency\footnote{Note, in this sense, that the Reissner-Nordstr\"om solution of GR represents an incomplete problem because the source term is generally not considered, restricting the discussion to the region external to the sources \cite{Ortin}.}  to the field equations of the static problem, in which the electric field is assumed sourceless, and demands a debate on the physical meaning and implications of curvature divergences since, as we have shown, they pose no obstacle to the existence of a well-defined (topological) electric flux through them.

Since the line element (\ref{eq:ds2static1}) recovers the Reissner-Nordstr\"{o}m geometry when $|x|\gg r_c^{(0)}$, one can verify that for  $\delta_1 = \delta_1^*$ an external horizon exists in general.  One thus expects that the existence of the horizon forces the regular configurations to decay into those with $\delta_1 \neq \delta_1^*$ via Hawking radiation. However, as shown in \cite{or12a,lor13}, when the number of charges $N_q\equiv q/|e|$ drops below the critical value  $N_q^c=\sqrt{2/\alpha_{em}} \approx 16.55$ (where $\alpha_{em}$ is the fine structure constant) the event horizon disappears, yielding an object which is stable against Hawking decay and whose charge is conserved and protected on topological grounds. Such everywhere regular and horizonless objects can be connected with black hole states, which posses an event horizon, in a continuous way, thus suggesting that they can be interpreted as black hole remnants.

As a final remark, we point out that for arbitrary $\delta_1$ the spatial integration of the action, representing the addition of electromagnetic plus gravitational energies, yields a finite result, which implies that the total energy is finite regardless of the existence or not of curvature divergences at the wormhole throat. In the particular case of the regular solutions $\delta_1 = \delta_1^*$, the action defined by Eqs. (\ref{eq:action}) and (\ref{eq:grav-lagrangian}) evaluated on the solutions  coincides with the action of a point-like massive particle at rest. Additionally, the surface $r=r_c^{(0)}$ becomes timelike when $N_q<N_q^c$, which further supports the idea that such regular solutions possess particle-like properties \cite{or12a,lor13}, representing a specific realization of Wheeler's geon \cite{Wheeler}.

\subsection{Dynamical charged configurations}

When the incoming null flux of radiation carries electric charge, the geometry changes in a highly non-trivial way. This setup should provide a good description of highly relativistic charged particles collapsing in a spherically symmetric way.  To illustrate this complexity, consider first that the initial state is  flat Minkowski space. Assume that a charged perturbation of compact support propagates  within the interval $[v_i,v_f]$.  Given the relation (\ref{eq:rtx0}), which for future reference we write as
\begin{equation}\label{eq:rtx1}
r^2(x,v)=\frac{x^2+\sqrt{x^4+4r_c^4(v)}}{2} \ ,
\end{equation}
where $r_c^4 (v)\equiv l_P^2 \tilde{\kappa}^2q^2(v)$, it follows that for $v<v_i$ the radial function $r^2(x,v)=x^2$ extends from zero to infinity \cite{lmor14}. As we get into the $v\ge v_i$ region, this radial function, which measures the area of the $2$-spheres of constant $x$ and $v$, never becomes smaller than $r_c^2 (v)$. In the region $v>v_f$, in which the ingoing flux of charge and radiation is again zero, the result is a static geometry identical to that described above in  Sec. \ref{sec:static}. This change in the geometry can be interpreted as the formation of a wormhole whose throat has an area $A_{WH}=4\pi r_c^2(v_f)$.

Depending on the total amounts of charge and energy conveyed by the incoming flux, the space-time may  have developed event horizons (see  Sec. \ref{sec:static}, and \cite{or12a} for full details on the different configurations). The existence or not of curvature divergences at $r=r_c(v_f)$ depends on the (integrated) charge-to-mass ratio of the flux. For simplicity, one can assume situations where $\delta_1(v_f)=\delta_1^*$, for which the final configuration has no curvature divergences, and $\delta_2(v_f)>\delta_1^*$, for which there are no event horizons. Related to this, we emphasize that, as shown  in Sec. \ref{sec:static}, the electric flux $\Phi$ through any $2$-surface enclosing the region $r=r_c(v)$ is always well-defined regardless of the value of $\delta_1(v_f)$.
\begin{figure}[h]
\includegraphics[width=0.35\textwidth]{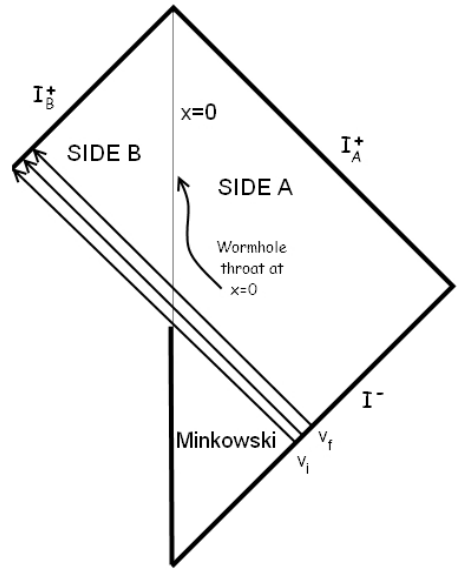}
\caption{Penrose diagram for the nonsingular case $\delta_1=\delta_1^*$ without event horizon, $N_q<N_q^c$.
Consider first that the initial state is flat Minkowski space. Next, assume that a charged perturbation of compact support propagates  within the interval $[v_i,v_f]$. We verify that the area of the $2$-spheres of constant $x$ and $v$ increases and never becomes smaller than $r_c^2 (v)$. If we consider now the region $v>v_f$, where the ingoing flux of charge and radiation is switched off, the result is a static geometry identical to that described in Section \ref{sec:static}. This change in the geometry can be interpreted as the formation of a wormhole whose throat has an area $A_{WH}=4\pi r_c^2(v_f)$. See the text for more details.
\label{fig:WH}}
\end{figure}

\begin{figure}[h]
\includegraphics[width=0.45\textwidth]{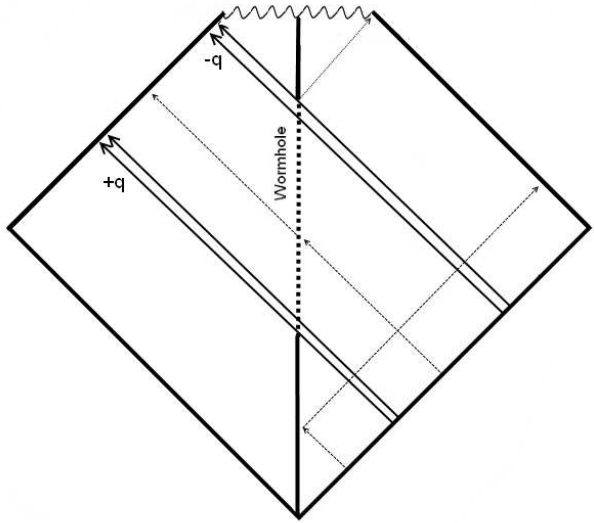}
\caption{Penrose diagram for the formation of a wormhole (with $\delta_1=\delta_1^*$ and $N_q<N_q^c$) out of Minkowski space resulting from the perturbation of a charged null fluid of integrated charge $+q$, and its subsequent removal due to a second flux of integrated charge $-q$. Note that we have chosen positive energy fluxes in both cases, which implies that the final state is a Schwarzschild black hole instead of Minkowski space.
\label{fig:WH2fluxes}}
\end{figure}

In a first approximation, this process of wormhole formation could be visualized as depicted in Fig. \ref{fig:WH}. This diagram suggests that the ingoing charged flux of radiation generates a whole new region of space-time as it propagates. This view would imply a change in the global  properties of space-time and, therefore, in its topology. A more careful examination of this process is necessary to understand how the {\it other side} of the wormhole arises and how this affects the topology of the problem. In fact, from a mathematical point of view, the exactly Minkowskian case $q=0$ can be seen as an exceptional situation in which the derivative of $r^2=x^2$ takes the values $\pm 1$ and has a  discontinuity at $x=0$. However, in a physical context with continuous virtual pair creation/annihilation out of the quantum vacuum, it seems  reasonable to expect that the exact case $q=0$ is never physically realized and that only the limiting case $q\to 0$ makes sense\footnote{ In such a scenario,  the vanishing of (the quantum average) $<q>$ in a given region can still be compatible with $<q^2>\neq 0$. In this sense, we understand that it is $<q^2>$ which should enter in the definition of $r_q^4(v)=l_P^2\tilde{\kappa}^2 <q^2(v)>$.}.
One can thus assume that in the physical branch of the theory, $q$ can be arbitrarily small but non-zero, with the limit $q\to 0$ leading to vanishing derivative $dr/dx$ at $x=0$ and quickly converging to $\pm 1$ away from $x=0$. In this scenario, the initial $q\to 0$ configuration could be seen as consisting of two identical pieces of Minkowski space-time connected along the line $x=0$ (see Fig. \ref{fig:WH2fluxes})  through a wormhole of area $A_{WH}\propto q \to 0$, which can be as small as one wishes but never zero due to the vacuum fluctuations.

We can now consider again the collapse of a spherical shell of charged radiation. As shown in Fig. \ref{fig:WH2fluxes}, the geometry inside the collapsing shell is essentially Minkowskian, up to the existence of infinitesimally small wormholes generated by quantum fluctuations (which realize the idea of a space-time foam). Though the details of the transient are complex and require a case-by-case numerical analysis because of the  $\rho_{in}$ term appearing in the line element (\ref{eq:ds2final}), the result of the collapse is the stretching of an initially infinitesimal wormhole to yield a  finite size hole of area $A_{WH}(v)=4\pi r_c^2(v)$. This occurs in such a way that the density of lines of force at the wormhole throat is kept constant, $\Phi (v)/A_{WH}(v)=\sqrt{c^7/(2\hbar G^2)}$. Note that this constraint between the flux and the area of the wormhole is valid for arbitrary charge and, in particular, in the limit $q\to 0$. Only if $q=0$ exactly, this ratio becomes indefinite. In GR, where an electric flux is assumed to be generated by a point-like particle (of zero area), one finds a divergent result. This divergence corresponds to taking the limit $\hbar\to 0$ in the above ratio and indicates that the wormhole closes in the limit in which classical GR is recovered, which is fully consistent with the fact that wormholes supported by electromagnetic fields do not exist in the case of GR \cite{Arellano}.

The sudden generation of a new space-time region depicted in Fig. (\ref{fig:WH}) can thus be avoided by assuming from the very beginning that the space-time admits a foam-like microstructure in which electric field lines may sustain wormholes that connect two different regions (the two sides of each wormhole).
The spontaneous generation of virtual pairs of electrically charged particles could be seen as the spontaneous formation of two nearby wormholes with identical but opposite charges \cite{Lobo:2014fma}. The energy deficit resulting from the generation of this pair would be released when the pair meets and the wormholes disappear. In this picture, the universe that we perceive would thus be a copy of another universe containing the same particles but with opposite charges (due to the different orientation of the fluxes on each side of the wormholes).

If a second flux of charged radiation is considered, the wormhole can be reduced again to an infinitesimal structure (associated to quantum fluctuations) if  at $v\ge v_{f_2}$ we have $q\to 0$ (see Fig. \ref{fig:WH2fluxes}). The geometry then becomes essentially identical to that of a Schwarzschild black hole on both sides, with curvature scalars diverging at $x=0$. If a new flux of charged matter reaches $x=0$, then the wormhole throat should grow again to give consistency to the electromagnetic field equations and conservation laws.

\section{Summary and Discussion} \label{secVI}

We have worked out a simplified scenario of gravitational collapse in which new gravitational physics at high energies is introduced by means of quadratic curvature corrections in the gravitational Lagrangian. We have made use of two elements that simplify the mathematical analysis, namely, 1) spherical symmetry, and 2) a pressureless fluid. These simplifications have been traditionally used in theoretical discussions about gravitational collapse and the study of the properties of singularities. Obviously, neither 1) nor 2) can be exactly realized in nature but, nonetheless, they are very useful for theoretical analysis of the type considered here. Note, in this sense, that already in the first models of gravitational collapse worked out by Oppenheimer and Snyder \cite{FN-S}, the internal pressures of the collapsing fluids were neglected as it was understood that, above a certain threshold, rather than helping to prevent the collapse they contribute to increase the energy density, which further accelerates the process. Similarly, in the case of electrically charged black holes (the well-known Reissner-Nordstr\"om solution), for instance, the repulsive electric force of the collapsed matter does not help to alleviate the strength of the central singularity. Rather, the squared of the Riemann tensor increases its degree of divergence, going from ${R^\alpha}_{\beta\mu\nu}{R_\alpha}^{\beta\mu\nu}\sim 1/r^4$ in the Schwarzschild case to  ${R^\alpha}_{\beta\mu\nu}{R_\alpha}^{\beta\mu\nu}\sim 1/r^8$ in the charged case. The energy of the electric field, therefore, worsens the degree of divergence of the curvature scalars.

In our model, we have considered a radiation fluid carrying a certain amount of energy and also electric charge. The repulsive forces or pressures that the particles making up the fluid could feel have been neglected as they are not essential for the study of the end state of the collapse. As a result, the fluid follows geodesics of the metric, which have been determined dynamically by taking into account the energy and charge conveyed by the fluid. The fluid motion, therefore, is not given a priori, but follows from the consistent resolution of the coupled system of radiation, electric field, and gravity.

The key point of this work has been the study of the end state of the collapse of this idealized system. In general relativity, this configuration unavoidably leads to the formation of a point-like singularity. In our model, however, the geometry and the topology undergo important changes. When the energy density of the collapsing fluid reaches a certain scale (of order the Planck scale), gravity is no longer attractive and becomes repulsive. This has a dramatic effect on the geodesics followed by the fluid which, rather than focusing into a point-like singularity, expand into a growing sphere. The wormhole is thus somehow produced by the repulsive character of gravitation at high energy-densities and the need to conserve the electric flux.

In principle, in our model wormholes of arbitrary charge and mass can be formed. However, this cannot be completely true since our results are valid as long as the approximations involved hold with sufficient accuracy. Therefore, one should note that in low-energy scenarios pressure and other dispersion effects should act so as to prevent the effective concentration of charge and energy way before it can concentrate at Planckian scales, thus suppressing wormhole production. However, for adequate concentrations of charge and energy, gravitational collapse cannot be halted and our analysis should be regarded as a good approximation. In this sense, we note that Hawking already analyzed the process of classical collapse in the early universe, finding that (primordial) black holes with a Planck mass or higher and up to 30 units of charge could be formed out of a charged plasma \cite{Hawking:1971ei}. Stellar collapse offers another robust mechanism to generate the conditions under which our approximations are valid. In fact, in order to build a completely regular configuration with a solar mass, about $\sim 10^{57}$ protons, one needs $\sim 2.91 \times 10^{21}$ electrons \cite{or12a}, which is a tiny fraction of the total available charge ($10^{-31}$) and mass. Therefore, the generation of wormholes under realistic situations is possible.

The dynamical generation of wormholes outlined above, in the context of charged fluids in quadratic Palatini gravity, differs radically in nature to the construction of general relativistic traversable wormholes, with the idealization of impulsive phantom radiation considered extensively in the literature \cite{Gergely02,Hayward02, Hayward:2001ma,Hayward04,Koyama:2002nh,Koyama:2004uh,Shinkai:2002gv,Hayward:2009yw}. In the latter, it was shown that two opposing streams of phantom radiation, which form an infinitely thin null shell, may support a static traversable wormhole \cite{Hayward02}. Essentially, one begins with a Schwarzschild black hole region, and triggers off beams of impulsive phantom radiation, with constant energy density profiles, from both sides symmetrically, consequently forming Vaidya regions. Now, in principle, if the energies and the emission timing are adequately synchronized, the regions left behind the receding impulses after the collision results in a static traversable wormhole geometry. Furthermore, it is interesting to note that it was shown that with a manipulation of the impulsive beams, it is possible to enlarge the traversable wormhole (see \cite{Hayward04} for more details). These solutions differ radically from the self-inflating wormholes discovered numerically \cite{Shinkai:2002gv} and the possibility that inflation might provide a natural mechanism for the enlargement of Planck-size wormholes to macroscopic size \cite{Roman:1992xj}. The difference lies in the fact that the amount of enlargement can be controlled by the amount of energy or the timing of the impulses, so that a reduction of the wormhole size is also possible by reversing the process of positive-energy and negative-energy impulses outlined in \cite{Hayward04}.

The theory presented here allows to generate static wormholes by means of a finite pulse of charged radiation, without the need to keep two energy streams active continuously or to synchronize them in any way across the wormhole. Regarding the size of the wormholes, we note that if instead of using $l_P^2$ to characterize the curvature corrections one considers a different length scale, say $l_\epsilon^2$, then their area would be given by $A_{WH}=\left(\frac{l_\epsilon}{l_P}\right)\frac{2N_q}{N_q^c}A_P$, where $A_P=4\pi l_P^2$, $N_q=|q/e|$ is the number of charges, and $N_q^c\approx 16.55$. Though this could allow to reach sizes orders of magnitude larger than the Planck scale, it does not seem very likely that macroscopic wormholes could arise from any viable theory of this form, though the role that other matter/energy sources could produce might be nontrivial.

Relative to the issue of classical singularities, the meaning and implications of the latter has been a subject of intense debate in the literature for years. Their existence in GR is generally interpreted as a signal of the limits of the theory, where quantum effects should become relevant and an improved theory would be necessary. This is, in fact, the reason that motivates our heuristic study of quadratic corrections beyond GR. As pointed out above and shown in detail in \cite{or12a}, the curvature divergences for the static wormhole solutions arising in quadratic Palatini gravity with electrovacuum fields (and also in the Palatini version of the Eddington-inspired Born-Infeld theory of gravity, see \cite{orh14}) are much weaker than their counterparts in GR (from $\sim 1/r^8$ in GR to $\sim 1/(r-r_c)^3$ in our model). Additionally, the existence of a wormhole structure that prevents the function $r^2$ from  dropping below the scale $r_c^2$ implies that the total energy stored in the electric field is finite (see \cite{orh14,lor13} for details), which clearly contrasts with the infinite result that GR yields. Therefore, even though curvature scalars may diverge, physical magnitudes such as total mass-energy, electric charge, and density of lines of force are insensitive to those divergences, which demands for an in-depth analysis of their meaning and implications. In this sense, we note that topology is a more primitive concept than geometry, in the sense that the former can exist without the latter. Comparison between a sphere and a cube is thus pertinent and enlightening in this context to better understand the physical significance of curvature divergences. It turns out that a cube and a sphere are topological equivalent. However, the geometry of the former is ill-defined along its edges and vertices. The divergent behavior of curvature scalars for certain values of $\delta_1$, therefore, simply indicates that for those cases the geometry is not smooth enough at the wormhole throat, but that does not have any impact on the physical existence of the wormhole.

Regarding the existence of curvature divergences at $x=0$ in the Schwarzschild case ($q\to 0$), our view is that there exist reasons to believe that such divergences could be an artifact of the approximations and symmetries involved in our analysis. These suspects are supported by the fact that radiation fluids (with equation of state $P/\rho=1/3$) in cosmological scenarios governed by the dynamics of the theory under study are able to avoid the Big Bang singularity, which is replaced by a cosmic bounce \cite{Barragan2010}. For the radiation fluid, the cosmic bounce occurs in both isotropic and anisotropic homogeneous scenarios when the energy density approaches the Planck scale. One would thus expect that a process of collapse mimicking the Oppenheimer-Snyder model with a radiation fluid should avoid the development of curvature divergences. This, in fact, occurs in Eddington-inspired Born-Infeld gravity \cite{orh14}, studied recently in \cite{Panietal}. The generic existence of  curvature divergences in the uncharged case involving a Vaidya-type scenario with null fluids is thus likely to be due to the impossibility of normalizing the null fluid, which is therefore insensitive to the existence of a limiting density scale. The consideration of more realistic non-null charged fluids could thus help to improve the current picture and avoid the shortcomings of the uncharged ($q\to 0$) Schwarzschild configurations.

As a final comment, we note that since in our theory the field equations outside the matter sources recover those of vacuum GR, Birkhoff's theorem must hold in those regions. This means that for $v<v_i$ we have Minkowski space, whereas for $v>v_f$ we have a Reissner-Nordstr\"om-like geometry of the form (\ref{eq:ds2static1}). The departure from Reissner-Nordstr\"om is due to the Planck scale corrections of the Lagrangian, which are excited by the presence of an electric field, and only affect the microscopic structure, which is of order $\sim r_c(v)$ (see Sec. \ref{sec:static} and \cite{or12a}).  Due to the spherical symmetry and the second-order character of the field equations, Birkhoff's theorem guarantees the staticity of the solutions for $v>v_f$.

To conclude, in this work an exact analytical solution for the dynamical process of collapse of a null fluid carrying energy and electric charge has been found in a quadratic extension of GR formulated \`{a} la Palatini. This scenario extends the well-known Vaidya-Bonnor solution of GR \cite{BV}, thus allowing to explore in detail new physics at the Planck scale. In the context of the static configurations, we have shown that wormholes can be formed out of Minkowski space by means of a pulse of charged radiation, which contrasts with previous approaches in the literature requiring artificial configurations and synchronization of two streams of phantom energy. Our results support the view that space-time could have a foam-like microstructure with wormholes generated by quantum fluctuations. Though such geometric structures develop, in general, curvature divergences, they are characterized by well-defined and finite electric charge and total energy. The physical role that such divergences could have is thus uncertain and requires an in-depth analysis, though from a topological perspective they seem not to play a relevant role. To fully understand these issues our model should be improved to address several important aspects including, for instance, the presence of gauge field degrees of freedom, to take into account the dynamics of counter-streaming effects due to the presence of simultaneous ingoing and outgoing fluxes, or to consider other theories of gravity beyond the quadratic Lagrangian (\ref{eq:grav-lagrangian}). These and related research issues are currently underway.

\section*{Acknowledgments}

F.S.N.L. acknowledges financial support of the Funda\c{c}\~{a}o para a Ci\^{e}ncia e Tecnologia through an Investigador FCT Research contract, with reference IF/00859/2012, funded by FCT/MCTES (Portugal), and grants CERN/FP/123615/2011 and CERN/FP/123618/2011. G.J.O. is supported by the Spanish grant FIS2011-29813-C02-02, the Consolider Program CPANPHY-
1205388, the JAE-doc program of the Spanish Research Council (CSIC), and the i-LINK0780 grant of CSIC. D.R.-G. is supported by CNPq (Brazilian agency) through project No. 561069/2010-7 and acknowledges hospitality and partial support from the Department of Physics of the University of Valencia, where this work initiated. This work has also been supported by CNPq project No. 301137/2014-5.


\begin{thebibliography}{99}


\bibitem{Vaidya}
P. C. Vaidya, Proc. Indian Acad. Sci. A \textbf{33}, 264 (1951); reprinted Gen. Rel. Grav. \textbf{31}, 119 (1999).

\bibitem{BV} W. B. Bonnor and P. C. Vaidya, Gen. Rel. Grav. \textbf{1}, 127 (1970).

\bibitem{Lake92} K. Lake, Phys. Rev. Lett. \textbf{68}, 3129 (1992);
P. S. Joshi, \emph{Global aspects in Gravitation and Cosmology} (Oxford University Press, 1993);
R. -G. Cai and A. Wang, Phys. Rev. D \textbf{73}, 063005 (2006).

\bibitem{radiation} W. A. Hiscock, Phys. Rev. D \textbf{23}, 2813 (1981); R. Parentani, Phys. Rev. D \textbf{63}, 041503(R) (2001).

\bibitem{Lasky07} P. D. Lasky and A. W. C. Lun, Phys. Rev. D \textbf{75}, 104010 (2007).

\bibitem{theor} A. K. Dawood and S. G. Ghosh, Phys. Rev. D \textbf{70}, 104010 (2004);
S. G. Ghosh and D. Kothawala, Gen. Rel. Grav. \textbf{40}, 9 (2008).

\bibitem{Ghosh12a} S. G. Ghosh and S. D. Maharaj, Phys. Rev. D \textbf{85}, 124064 (2012).

\bibitem{Cai08} R. -G. Cai, L. -M. Cao, Y. -P. Hu, and S. P. Kim, Phys. Rev. D \textbf{78}, 124012 (2008).

\bibitem{Hayward02}
S. A. Hayward, Phys. Rev. D \textbf{65}, 124016 (2002).

\bibitem{Gergely02} L. A. Gergely, Phys. Rev. D \textbf{65}, 127503 (2002); \textbf{58}, 084030 (1998).

\bibitem{Hayward:2001ma}
  S.~A.~Hayward, S.~-W.~Kim, and H.~-J.~Lee,
  %``Dilatonic wormholes: Construction, operation, maintenance and collapse to black holes,''
  Phys.\ Rev.\ D {\bf 65}, 064003 (2002);
  %[gr-qc/0110080].
  %%CITATION = GR-QC/0110080;%%
S.~A.~Hayward,
  %``Dynamic wormholes,''
  Int.\ J.\ Mod.\ Phys.\ D {\bf 8}, 373 (1999).
  %[gr-qc/9805019].
  %%CITATION = GR-QC/9805019;%%

\bibitem{Hayward04}
S.~A.~Hayward and H.~Koyama,
  %``How to make a traversable wormhole from a Schwarzschild black hole,''
  Phys.\ Rev.\ D {\bf 70}, 101502 (2004).
  %[gr-qc/0406080].
  %%CITATION = GR-QC/0406080;%%


\bibitem{Planck2} P. A. R. Ade et al., Planck 2013 results. XVI, arXiv: 1303.5076
[astro-ph] (2013).

\bibitem{Morris}
M. Morris and K. S. Thorne,
%``Wormholes in space-time and their use
%for interstellar travel: A tool for teaching General Relativity,''
Am. J. Phys. \textbf{56}, 395 (1988).

\bibitem{phantomWH}
S.~V.~Sushkov,
  %``Wormholes supported by a phantom energy,''
  Phys.\ Rev.\ D {\bf 71}, 043520 (2005);
  %[gr-qc/0502084].
  %%CITATION = GR-QC/0502084;%%
  F.~S.~N.~Lobo,
  %``Phantom energy traversable wormholes,''
  Phys.\ Rev.\ D {\bf 71}, 084011 (2005);
  %[gr-qc/0502099].
  %%CITATION = GR-QC/0502099;%%
  %``Stability of phantom wormholes,''
  {\bf 71}, 124022 (2005).
  %[gr-qc/0506001].
  %%CITATION = GR-QC/0506001;%%

\bibitem{Wheeler}
J. A. Wheeler, Phys. Rev. \textbf{97}, 511 (1955).

\bibitem{Roman:1992xj}
  T.~A.~Roman,
  %``Inflating Lorentzian wormholes,''
  Phys.\ Rev.\ D {\bf 47}, 1370 (1993).
  %[gr-qc/9211012].
  %%CITATION = GR-QC/9211012;%%

\bibitem{Hayward:2004wm}
  S.~A.~Hayward and H.~Koyama,
  %``How to make a traversable wormhole from a Schwarzschild black hole,''
  Phys.\ Rev.\ D {\bf 70}, 101502 (2004).
  %[gr-qc/0406080].
  %%CITATION = GR-QC/0406080;%%

\bibitem{Lobo:2005uf}
  F.~S.~N.~Lobo,
  %``Stable dark energy stars,''
  Class.\ Quant.\ Grav.\  {\bf 23}, 1525 (2006).
  %[gr-qc/0508115].
  %%CITATION = GR-QC/0508115;%%

\bibitem{DeBenedictis:2008qm}
  A.~DeBenedictis, R.~Garattini, and F.~S.~N.~Lobo,
  %``Phantom stars and topology change,''
  Phys.\ Rev.\ D {\bf 78}, 104003 (2008).
  %[arXiv:0808.0839 [gr-qc]].
  %%CITATION = ARXIV:0808.0839;%%

\bibitem{Redmount:1992mc}
  I.~H.~Redmount and W.~-M.~Suen,
  %``Is quantum space-time foam unstable?,''
  Phys.\ Rev.\ D {\bf 47}, 2163 (1993);
  %[gr-qc/9210017].
  %%CITATION = GR-QC/9210017;%%
  %``Quantum dynamics of Lorentzian space-time foam,''
  {\bf 49}, 5199 (1994).
  %[gr-qc/9309017].
  %%CITATION = GR-QC/9309017;%%

\bibitem{acausal}
R. Geroch, J. Math. Phys. \textbf{8}, 782 (1967);
S. W. Hawking, Phys. Rev. D  \textbf{46}, 603 (1992).

\bibitem {Visser}
M. Visser, \textit{Lorentzian Wormholes: From Einstein to
Hawking} (American Institute of Physics, New York, 1995).

\bibitem {dewitt}
A.~Anderson and B.~S.~DeWitt,
  %``Does The Topology Of Space Fluctuate?,''
  Found.\ Phys.\  {\bf 16}, 91 (1986).
  %%CITATION = FNDPA,16,91;%%

\bibitem{Visser:1989ef}
  M.~Visser,
  %``Wormholes, Baby Universes And Causality,''
  Phys.\ Rev.\ D {\bf 41}, 1116 (1990);
  %%CITATION = PHRVA,D41,1116;%%
  %
%M.~Visser,
  %``Quantum wormholes,''
  %Phys.\ Rev.\ D
  {\bf 43}, 402 (1991).
  %%CITATION = PHRVA,D43,402;%%


\bibitem{Garattini:2013pha}
  R.~Garattini and F.~S.~N.~Lobo,
  %``Gravity's Rainbow induces Topology Change,''
  arXiv:1303.5566 [gr-qc].
  %%CITATION = ARXIV:1303.5566;%%

\bibitem{Garattini}
  R.~Garattini and F.~S.~N.~Lobo,
  %``Self sustained phantom wormholes in semi-classical gravity,''
  Class.\ Quant.\ Grav.\  {\bf 24}, 2401 (2007);
  %[gr-qc/0701020].
  %%CITATION = GR-QC/0701020;%%
%R.~Garattini and F.~S.~N.~Lobo,
  %``Self-sustained traversable wormholes in noncommutative geometry,''
  Phys.\ Lett.\ B {\bf 671}, 146 (2009);
  %[arXiv:0811.0919 [gr-qc]].
  %%CITATION = ARXIV:0811.0919;%%
%R.~Garattini and F.~S.~N.~Lobo,
  %``Self-sustained wormholes in modified dispersion relations,''
  Phys.\ Rev.\ D {\bf 85}, 024043 (2012).
  %[arXiv:1111.5729 [gr-qc]].
  %%CITATION = ARXIV:1111.5729;%%

\bibitem{geons4b}
P. R. Anderson and D. R. Brill,
%``Gravitational geons revisited,''
Phys. Rev. D \textbf{56}, 4824 (1997).

\bibitem{or12a}  G. J. Olmo and D. Rubiera-Garcia, Phys. Rev. D \textbf{86}, 044014 (2012);
Int. J. Mod. Phys. D \textbf{21}, 1250067 (2012);
Eur. Phys. J. C \textbf{72}, 2098 (2012).

\bibitem{lor13}
F.~S.~N.~Lobo, G.~J.~Olmo, and D.~Rubiera-Garcia,
  %``Semiclassical geons as solitonic black hole remnants,''
  JCAP {\bf 07}, 011 (2013).
  %[arXiv:1306.2504 [hep-th]].
  %%CITATION = ARXIV:1306.2504;%%

\bibitem{lmor14}  F. S. N. Lobo, J. Martinez-Asencio, G. J. Olmo, and D. Rubiera-Garcia, Phys. Lett. B, in press (2014).

\bibitem{Zanelli} J. Zanelli, arXiv:hep-th/0502193.

\bibitem{or13a}
  G.~J.~Olmo and D.~Rubiera-Garcia, Phys. Rev. D \textbf{88}, 084030 (2013).
  %``Importance of torsion and invariant volumes in Palatini theories of gravity,''
  %arXiv:1306.4210 [hep-th].

\bibitem{Barragan2010} C. Barragan and G. J. Olmo, Phys. Rev. D \textbf{82}, 084015 (2010);
% Isotropic and Anisotropic Bouncing Cosmologies in Palatini Gravity
% e-Print: arXiv:1005.4136 [gr-qc]
C. Barragan, G. J. Olmo, and H. Sanchis-Alepuz, Phys. Rev. D \textbf{80}, 024016 (2009).
% Bouncing Cosmologies in Palatini f(R) Gravity
% e-Print: arXiv:0907.0318 [gr-qc]

\bibitem{Franca} M. Ferraris, M. Francaviglia, and I. Volovich, Class. Quant. Grav. \textbf{11}, 1505 (1994);
%[gr-qc/9303007]
%The Universality of vacuum Einstein equations with cosmological constant
A. Borowiec, M. Ferraris, M. Francaviglia, and I. Volovich, Class. Quant. Grav. \textbf{15}, 43 (1998).
%Universality of Einstein equations for the Ricci squared Lagrangians [gr-qc/9611067] [INSPIRE].

\bibitem{Stephani} H. Stephani et al., \emph{Exact solutions to Einstein's field equations} (Cambridge University Press, 2003).

\bibitem{PT} L. Parker and D. J. Toms, \emph{Quantum field theory in curved space-time: quantized fields and gravity} (Cambridge University Press U. K., 2009);
    N. D. Birrel and P. C. W. Davies, \emph{Quantum fields in curved space} (Cambridge University Press, U. K., 1982).

\bibitem{strings1} M. Green, J. Schwarz, and E. Witten, \emph{Superstring theory} (Cambridge University Press, U. K., 1987).
\bibitem{Ortin} T. Ortin, \emph{Gravity and strings} (Cambridge University Press, U. K. 2004).

\bibitem{Cembranos} J. A. R. Cembranos, Phys. Rev. Lett. \textbf{102}, 141301 (2009).

\bibitem{o08}
G.~J.~Olmo and P.~Singh,
  %``Effective Action for Loop Quantum Cosmology a la Palatini,''
  JCAP \textbf{0901}, 030 (2009).

\bibitem{lqc}
M. Bojowald, Living Rev. Rel. \textbf{11}, 4 (2008); G. A. Mena-Marugan, AIP Conf. Proc. \textbf{1130}, 89 (2009); J. Phys. Conf. Ser. \textbf{314}, 012012 (2011); A.~Ashtekar and P.~Singh,  Class.\ Quant.\ Grav.\  \textbf{28}, 213001 (2011).


\bibitem{Deser} S. Deser and G. W. Gibbons, Class. Quant. Grav. \textbf{15}, L35 (1998).

\bibitem{Banados} M. Ba\~nados and P. G. Ferreira, Phys. Rev. Lett. \textbf{105}, 011101 (2010).

\bibitem{orh14}  G. J. Olmo, D. Rubiera-Garcia, and H. Sanchis-Alepuz, Eur. Phys. J. C, in press (2014).

\bibitem{or13b} E. Guendelman, G. J. Olmo, D. Rubiera-Garcia, and M. Vasihoun, Phys. Lett. B \textbf{726}, 870 (2013).

\bibitem{or13c} G. J. Olmo and D. Rubiera-Garcia, JCAP \textbf{02}, 010 (2014).

\bibitem{or12d} J. Martinez-Asencio, G. J. Olmo, and D. Rubiera-Garcia, Phys. Rev. D \textbf{86}, 104010 (2012).

\bibitem{JMMG} J. M. Martin-Garcia, \url{http://www.xact.es.}

\bibitem{Misner:1957mt}
  C.~W.~Misner and J.~A.~Wheeler,
  %``Classical physics as geometry: Gravitation, electromagnetism, unquantized charge, and mass as properties of curved empty space,''
  Annals Phys.\  {\bf 2}, 525 (1957).
  %%CITATION = APNYA,2,525;%%

\bibitem{Arellano} A.~V.~B.~Arellano and F.~S.~N.~Lobo, Class. Quant. Grav. \textbf{23}, 7229 (2006); 5811 (2006).

\bibitem{Lobo:2014fma}
  F.~S.~N.~Lobo, G.~J.~Olmo and D.~Rubiera-Garcia,
  %``Microscopic wormholes and the geometry of entanglement,''
  arXiv:1402.5099 [hep-th].

\bibitem{Koyama:2002nh}
  H.~Koyama, S.~A.~Hayward, and S.~-W.~Kim,
  %``Construction and enlargement of dilatonic wormholes by impulsive radiation,''
  Phys.\ Rev.\ D {\bf 67}, 084008 (2003).
  %[gr-qc/0212106].
  %%CITATION = GR-QC/0212106;%%

\bibitem{Koyama:2004uh}
  H.~Koyama and S.~A.~Hayward,
  %``Construction and enlargement of traversable wormholes from Schwarzschild black holes,''
  Phys.\ Rev.\ D {\bf 70}, 084001 (2004).
  %[gr-qc/0406113].
  %%CITATION = GR-QC/0406113;%%


\bibitem{Shinkai:2002gv}
  H.~-A.~Shinkai and S.~A.~Hayward,
  %``Fate of the first traversible wormhole: Black hole collapse or inflationary expansion,''
  Phys.\ Rev.\ D {\bf 66}, 044005 (2002).
  %[gr-qc/0205041].
  %%CITATION = GR-QC/0205041;%%

\bibitem{Hayward:2009yw}
  S.~A.~Hayward,
  %``Wormhole dynamics in spherical symmetry,''
  Phys.\ Rev.\ D {\bf 79}, 124001 (2009).
  %[arXiv:0903.5438 [gr-qc]].
  %%CITATION = ARXIV:0903.5438;%%

\bibitem{FN-S}
A. Fabbri and J. Navarro-Salas, {\it Modeling Black Hole Evaporation} (ICP-World Scientific, London, England, 2005).

\bibitem{Hawking:1971ei}
  S.~Hawking,
  %``Gravitationally collapsed objects of very low mass,''
  Mon.\ Not.\ Roy.\ Astron.\ Soc.\  {\bf 152}, 75 (1971).

\bibitem{Panietal}
P. Pani, V. Cardoso, and T. Delsate, Phys. Rev. Lett. {\bf 107}, 031101 (2011); P. Pani, T. Delsate, and V. Cardoso, Phys. Rev. D {\bf 85}, 084020 (2012).


\end{thebibliography}
\end{document}